\documentstyle[preprint,aps,epsfig]{revtex}

\begin{document}

\draft
\preprint{
\vbox{
\hbox{ADP-98-22/T298}
\hbox{IU/NTC 98-04}
}}

\title{Evidence for Substantial Charge Symmetry Violation in 
Parton Distributions} 

\normalsize
\author{C.Boros$^1$, J.T.Londergan$^2$ and A.W.Thomas$^1$}
\address {$^1$ Department of Physics and Mathematical Physics,
                and Special Research Center for the
                Subatomic Structure of Matter,
                University of Adelaide,
                Adelaide 5005, Australia}
\address{$^2$ Department of Physics and Nuclear
            Theory Center, Indiana University,
            Bloomington, IN 47404, USA}

\date{\today}
\maketitle

\begin{abstract}  
In principle one can test the validity of charge symmetry for parton 
distributions by comparing structure functions measured in neutrino and  
charged lepton deep inelastic scattering. New experiments  
make such tests possible; they provide rather tight upper limits on 
parton charge symmetry violation [CSV] for intermediate Bjorken $x$, 
but appear to show evidence for CSV effects at small $x$. We examine 
two effects which might account for this experimental discrepancy: 
nuclear shadowing corrections for neutrinos, and strange quark 
contributions $s(x) \ne \bar{s}(x)$. We show that neither of these 
two corrections removes the experimental discrepancy  
between the structure functions.   
We are therefore forced to consider the possibility of a surprisingly 
large CSV effect in the nucleon sea quark
distributions.  
 
\end{abstract}

PACS: 13.60.Hb, 13.15.+g, 12.40.Vv, 11.40.Ha

\tightenlines 

\newpage 

In nuclear physics charge symmetry, which interchanges protons
and neutrons (simultaneously interchanging up and down quarks),  
is respected to a high degree of precision.  Most  
low-energy tests of charge symmetry find that it is good to at least
$1\%$ in reaction amplitudes \cite{Miller}. Therefore, charge symmetry 
is usually assumed to be valid in discussions of strong interactions.   
Currently all phenomenological analyses describe deep inelastic scattering 
[DIS] data using charge symmetric parton distributions.  
Until recently this assumption seemed to be 
justified, since the quantitative evidence which could be extracted
from high energy experiments, although not particularly precise, was
consistent with charge symmetric parton distributions 
\cite{Lon98}.  

Experimental verification of charge symmetry is difficult, partly 
because the relative charge symmetry violation (CSV) effects 
are expected to be small, requiring high precision experiments, 
and partly because CSV often mixes with parton flavor symmetry violation 
(FSV).  Recent experimental measurements by the NMC Collaboration 
\cite{NMCfsv},  demonstrating the violation of the Gottfried 
sum rule \cite{Gottfried}, have been widely interpreted as 
evidence for what is termed SU(2) FSV. The measurement of the  
ratio of Drell-Yan cross sections in proton-deuteron and 
proton-proton scattering, first by the NA51-Collaboration at CERN 
\cite{Na51} and more recently by the E866 experiment 
at FNAL \cite{E866}, also indicate substantial FSV. 
However, as pointed out by Ma \cite{Ma1}, both of these experiments 
could in principle be explained by sufficiently large CSV effects, 
even in the limit of exact flavor symmetry.  
In view of these ambiguities in the interpretation of current 
experimental data, it would be highly desirable to have experiments 
which separate CSV from FSV.  
A few experiments have been already proposed 
\cite{Tim1} and could be carried out in the near future. 

At the level of parton distributions charge symmetry implies the 
equivalence between up (down) 
quark distributions in the proton and down (up) quarks 
in the neutron.  We define 
charge symmetry violating distributions 
\begin{eqnarray} 
\delta u(x)& =&  u^p(x) -d^n(x) \nonumber\\ 
\delta d(x)& =&  d^p(x) -u^n(x), 
\end{eqnarray} 
where the superscripts $p$ and $n$ refer to 
quark distributions in the proton and neutron, respectively (quark
distributions without subscripts will refer to the proton). 
The relations for CSV in antiquark distributions 
are analogous.  Exact charge symmetry would require that the 
quantities $\delta u(x)$ and $\delta d(x)$ vanish.

In the quark-parton model the structure functions measured in 
neutrino, antineutrino  and charged lepton DIS on an iso-scalar 
target, $N_0$,   
are given in terms of the parton distribution functions and the 
charge symmetry violating distributions 
\cite{Lon98}   
\begin{eqnarray}
  F_2^{\nu N_0} (x,Q^2) &=& x[u(x)+ \bar u(x) +d(x) +\bar d(x)
     + 2 s(x) + 2 \bar c(x) -\delta u(x)-\delta \bar d(x)] \nonumber \\ 
  F_2^{\bar \nu N_0} (x,Q^2) &=& x[u(x)+ \bar u(x) +d(x) +\bar d(x)
     + 2 \bar s(x) + 2  c(x) -\delta d(x)-\delta \bar u(x)] \nonumber \\
  xF_3^{\nu N_0}(x,Q^2) &=& x[u(x) + d(x) -\bar u(x) - \bar d(x) 
       +2 s(x)-2 \bar c(x) -\delta u(x) +\delta\bar d(x)]   
      \nonumber\\ 
  xF_3^{\bar\nu N_0}(x,Q^2) &=& x[u(x) + d(x) -\bar u(x) - \bar d(x) 
       -2 \bar s(x)+2  c(x) -\delta d(x) +\delta\bar u(x)]  \nonumber\\ 
  F_2^{\ell N_0}(x,Q^2) & =& \frac{5}{18} x 
   [ u(x) + \bar u(x)
 +d(x) +\bar d(x) + \frac{2}{5} (s(x) + \bar s(x)) 
  +  \frac{8}{5}(c(x)+\bar c(x)) \nonumber\\ 
 & -  & \frac{4}{5} 
 (\delta d(x)+\delta \bar d(x)) - \frac{1}{5} ( \delta u(x)+\delta 
   \bar u(x))]
\label{eq2} 
\end{eqnarray}

By comparing structure functions measured in neutrino and charged lepton 
deep inelastic scattering [DIS] on isoscalar targets, it is possible to 
test parton charge symmetry.  An example is 
the ``charge ratio'',  which relates the neutrino 
structure function to the structure function 
measured in charged lepton DIS 
\begin{eqnarray}
 R_c (x) & \equiv  & \frac{F_2^{\mu N_0}(x)}{\frac{5}{18}
 F_2^{\nu N_0}(x) -x( s(x) +\bar s(x))/6} \nonumber\\ 
&\approx & 1 - \frac{s(x) -\bar s(x)}{\overline{Q}(x)} + 
 \frac{4\delta u(x) - \delta \bar u(x) - 4 \delta d(x) 
+\delta \bar d(x)}{5 \overline{Q}(x)}.   
\label{rc}
\end{eqnarray}
Here, we defined 
$\bar Q(x) \equiv \sum_{q=u,d,s} (q(x)+\bar q(x))-3(s(x)+\bar s(x))/5$, 
and we have expanded Eq.\ \ref{rc} to lowest order in small quantities. 
A deviation $R_c(x) \ne 1$, at any value of $x$, must arise either 
from CSV effects or from an inequality between strange and anti-strange
quark distributions.  

\begin{figure}
\epsfig{figure=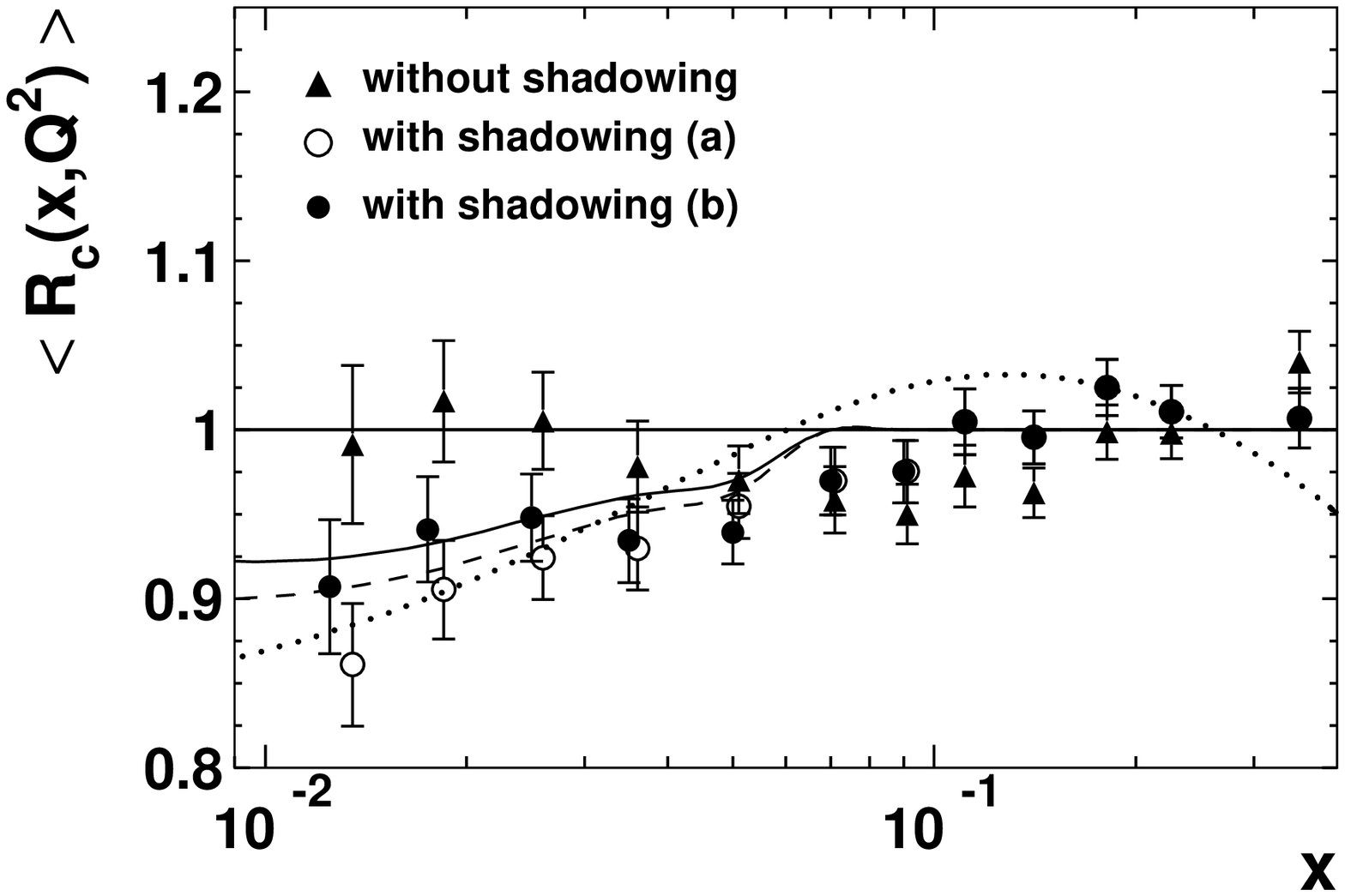,height=12.cm}
\caption{The ``charge ratio'' as a function of $x$  calculated using
         the CCFR \protect\cite{CCFR} data for neutrino and
         NMC \protect\cite{NMCfsv} data for muon
         structure functions. The data have been integrated
         over the overlapping kinematical regions
         and have been corrected for heavy target effects
         using a parametrization (dotted line) for heavy target corrections
         extracted from charged lepton scattering 
         (shadowing ``(a)''). The result is
         shown as open circles.
         The ratio  obtained without heavy target corrections
	 and with shadowing corrections calculated in the 
         ``two phase'' model (shadowing ``(b)'') 
          are  shown as solid triangles and circles, 
         respectively. 
         The calculated heavy target correction factors for neutrino and 
         for charged lepton scattering are represented by solid 
         and dashed lines, respectively.   
         Statistical and systematic errors are added in
         quadrature. }
\label{fig1}
\end{figure}

Recent experimental measurements make it possible
to carry out a precise comparison between $F^\nu_2(x,Q^2)$ 
and $F_2^{\mu}(x,Q^2)$.    
The CCFR-Collaboration compared the neutrino structure function
$F_2^\nu (x,Q^2)$ extracted from their data on an iron target \cite{CCFR}
with $F_2^\mu (x,Q^2)$ measured for the deuteron by the NMC
Collaboration \cite{NMC}.  In the region of intermediate values of
Bjorken $x$ ($0.1 \le x \le 0.4$), the two structure functions are 
in very good agreement; in this region we can set upper limits on 
parton CSV contributions of a few percent.  In the small
$x$-region however ($x < 0.1$), the CCFR group found that the two
structure functions differ by as much as 10-15$\%$. 
This can be seen in Fig.\ref{fig1} where the ``charge ratio'' 
has been obtained by integrating over the region of overlap in  
$Q^2$ of the two experiments.  The various data points in 
Fig.\ \ref{fig1} represent different ways of calculating 
nuclear shadowing corrections, as we will discuss.  
Several corrections must be applied to the data before any
conclusions may be drawn from this discrepancy. 
The CCFR Collaboration made a careful study of overall normalization, 
charm threshold and iso-scalar correction effects \cite{CCFR}. Here 
we discuss the most important remaining effects, heavy target 
corrections for the neutrinos and differences between strange 
and anti strange quark distributions.    

Nuclear shadowing corrections for neutrinos are generally accounted for
by using correction factors obtained from charged lepton reactions 
at the same kinematic values.  
{\it A priori}, there is no reason to assume   
that neutrino and charged lepton heavy target corrections should be 
identical.  
We re-examined heavy target corrections to deep-inelastic neutrino 
scattering, focusing on the differences between neutrino and charge 
lepton scattering and on effects due to the $Q^2$-dependence of 
shadowing for moderately large $Q^2$.  This work will be
published elsewhere \cite{Boros}.  We used a two phase model  
which has been successfully applied to the description of shadowing 
in charged lepton DIS  
\cite{Badelek}. In this approach,  
vector meson dominance was used to describe the low $Q^2$ virtual 
photon or $W$ interactions, and Pomeron exchange was used for the 
approximate scaling region.  In generalizing
this approach to weak currents, the major new features are that the 
axial-vector current is only partially conserved, and that the weak 
current couples to axial vector mesons in addition to vector mesons.  

Using this two phase model, we calculated the shadowing corrections 
to the CCFR neutrino data and used these corrections in 
calculating the charge ratio $R_c$ of 
Eq.\ \ref{rc}. The result is shown in 
Fig.\ref{fig1}.  The solid triangles show the charge ratio when no
shadowing corrections are used.  The open 
circles show the charge ratio when the neutrino data is modified 
using heavy target shadowing corrections from charged lepton reactions, 
and the solid circles show the results calculated specifically for
neutrinos using our two phase model.  For $x \ge 0.1$, the
two shadowing corrections give essentially identical results.  At 
small $x$, careful consideration of neutrino shadowing corrections 
decreases, but does not resolve, the low-$x$ discrepancy between the 
CCFR and NMC data. 

At this point it is instructive to review how the structure
functions are extracted in neutrino reactions.  Because of the
extremely small cross sections, it is necessary to integrate
cross sections over all energies in order to accumulate 
sufficient flux.  When this has been done, the resulting 
neutrino and antineutrino cross sections give two 
equations in the neutrino structure functions.    
If we assume that the neutrino and anti-neutrino 
structure functions are equal $F_2^\nu(x,Q^2)=F_2^{\bar\nu}(x,Q^2)$, 
with an analogous relation for $xF_3 (x,Q^2)$, then we have 
two linear equations in two unknowns.  From Eq.\ \ref{eq2} 
we see that these relations will be true if charge symmetry is
valid and if $s(x) = \bar{s}(x)$; there is an additional correction 
to $xF_3^\nu$ and $xF_3^{\bar\nu}$ from strange quarks.  

The resulting structure function $F_2^{CCFR}$ is a flux weighted average 
between neutrino and  
anti-neutrino structure functions (see Ref.\cite{CCFR}). 
This fact becomes important if charge symmetry is violated or 
the strange and anti-strange quark distributions are 
different.  If we define  
$\alpha = N_\nu /(N_\nu + N_{\bar\nu})$, where  
$N_\nu$ and $N_{\bar\nu}$ are the number of neutrino and anti-neutrino 
events, respectively, $F_2^{CCFR}(x,Q^2)$ is proportional to   
\begin{eqnarray}
  F_2^{CCFR}(x,Q^2)& = &\alpha F_2^\nu (x,Q^2) + (1-\alpha ) 
F_2^{\bar\nu}(x,Q^2)\nonumber\\ 
  &=& \frac{1}{2} [F_2^\nu(x,Q^2) + F_2^{\bar\nu}(x,Q^2)] 
+\frac{1}{2} (2\alpha -1) [F_2^\nu(x,Q^2) - F_2^{\bar\nu}(x,Q^2)].  
\label{s1}
\end{eqnarray}
This is equal to $\frac{1}{2}[ F_2^\nu(x,Q^2)+ 
F_2^{\bar\nu}(x,Q^2)]$ if $\alpha=\frac{1}{2}$ or if 
the two structure functions are equal 
(which implies  $s(x)=\bar s(x)$ and the validity of charge symmetry).   
The CCFR-Collaboration collected 
$1,300,000$ neutrino and $270,000$ anti-neutrino events 
\cite{CCFR}; thus $\alpha \approx 0.83$ so that to a good 
approximation $F_2^{CCFR}(x,Q^2)$ can be regarded as a neutrino 
structure function. 

The most likely explanation for the small-$x$ discrepancy in the 
charge ratio is either from different strange quark 
distributions $s(x) \ne \bar{s}(x)$ \cite{Melni97},    
or from charge symmetry violation. First, we examine   
the role played by the strange and anti-strange quark 
distributions.  Assuming charge symmetry,  
the strange and anti-strange quark distributions are given by a 
linear combination of the structure functions measured in neutrino 
and in muon DIS,   
\begin{equation} 
   \frac{5}{6} F_2^{CCFR}(x,Q^2) -3 F_2^{NMC}(x,Q^2)  
= \frac{1}{2}\, x \, [s(x) + \bar s(x)] 
+\frac{5}{6} (2\alpha -1)\, x \,[s(x)-\bar s(x)]. 
\label{diff} 
\end{equation} 
Under the assumption $s(x)=\bar s(x)$, this relation could be 
used to extract the strange quark distribution. 
However, as is well known, $s(x)$    
obtained in this way is inconsistent with the distribution 
extracted from independent experiments.  

The strange quark distribution can be determined directly 
from opposite sign dimuon production in deep inelastic neutrino and 
anti-neutrino scattering. The CCFR Collaboration performed a LO 
\cite{CCFRLO}  
and NLO analysis \cite{CCFRNLO} of their dimuon data using    
the neutrino (anti-neutrino) events to extract the strange 
(anti-strange) quark distributions. 
They found that $s(x)$ 
and $\bar{s}(x)$ were equal within experimental errors 
in NLO \cite{CCFRNLO}. However, since the number 
of anti-neutrino events is much smaller than that of the neutrino 
events, the errors of this analysis are inevitably large. 

Since the dimuon experiments are carried out on an iron target,  
shadowing corrections could also modify the extracted 
strange quark distribution.  
The CCFR-Collaboration normalized the dimuon cross section 
to the ``single muon'' cross section and argued that 
the heavy target correction should cancel in the ratio. 
However, the charm producing part of the structure function  
$F_2^{cp}(x,Q^2)$ could in principle be shadowed differently  
from the non-charm producing part $F_2^{ncp}(x,Q^2)$.  
We tested this hypothesis by calculating the neutrino shadowing 
corrections for both the 
charm and non-charm producing part of the structure 
function.  The results will be published separately \cite{Bor98}. 
We find that, while the relative importance 
of the Pomeron and VMD components are 
different in the charm producing ($cp$) and the non-charm producing 
($ncp$) parts, there is essentially no difference 
in the total shadowing. 

It appears plausible that the low-$x$ discrepancy 
in the charge ratio of Eq.\ \ref{rc} can be accounted for by allowing
$s(x) \ne \bar{s}(x)$. 
To test this hypothesis we combined the 
dimuon production data, averaged over $\nu$ and 
$\bar\nu$ events, with the structure functions from neutrino and 
charged lepton scattering (Eq.(\ref{diff})).   
Defining $\alpha^\prime= N_\nu/(N_\nu+N_{\bar\nu})$, where 
$N_\nu =5,030 $, $N_{\bar\nu}=1,060$ ($\alpha^\prime \approx 0.83$)  
are respectively the $\nu$ and $\bar\nu$ 
events from the dimuon production experiment \cite{CCFRNLO}, 
the flux-weighted experimental distribution $x s(x)^{\mu\mu}$ from 
dimuon production is   
\begin{equation} 
 x s^{\mu\mu}(x) = \frac{1}{2}\, x \,[s(x) + \bar s(x)] + \frac{1}{2}  
       (2\alpha^\prime -1 )\, x\, [s(x) - \bar s(x)].  
\label{s2}
\end{equation} 
This equation together with Eq.(\ref{diff}) 
forms a pair of linear equations which can be solved for 
$s(x)$ and  $\bar s(x)$.   
We can simultaneously test the compatibility of the various  
experiments. 

\begin{figure}
\epsfig{figure=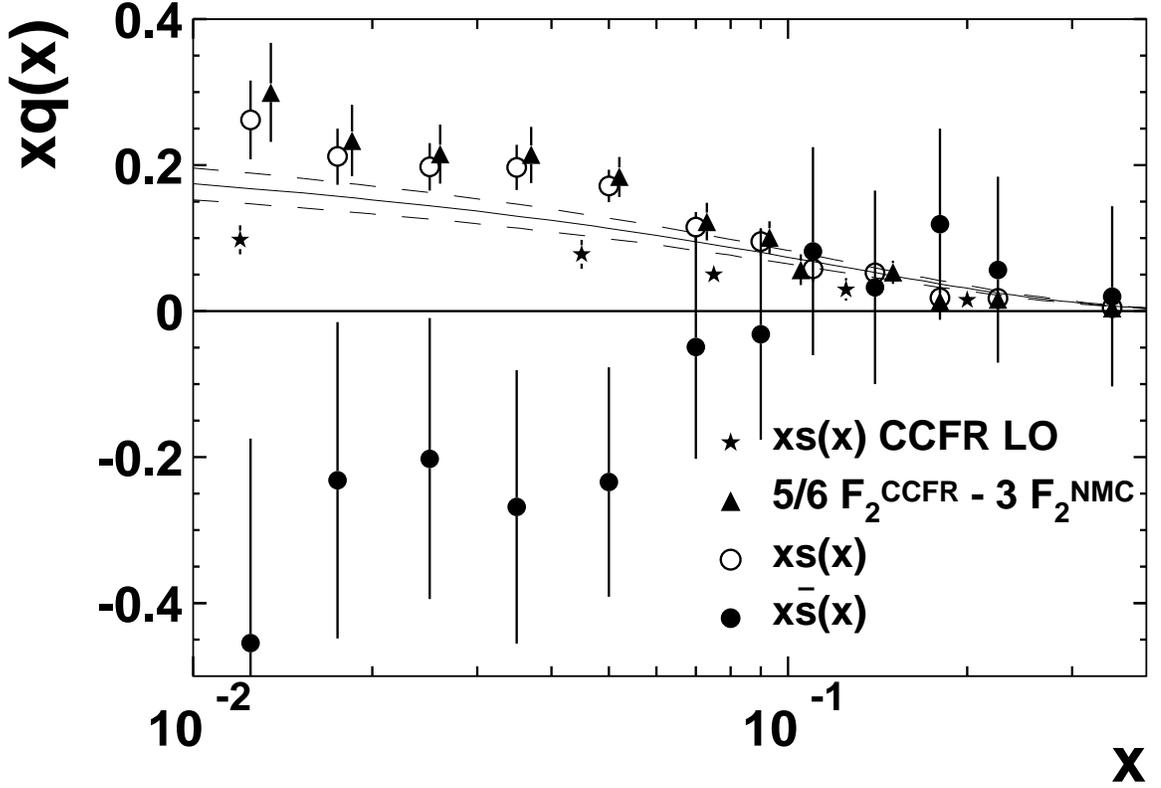,height=14.cm}
\caption{The strange quark distribution $s(x)$ (open circles) and 
      anti-strange distribution $\bar{s}(x)$ (solid circles) 
      extracted by combining the CCFR and NMC structure functions 
      with $s(x)$ extracted from dimuon experiments, as given in
      Eqs.\ \protect\ref{diff} and \protect\ref{s2}.  
      The difference between the CCFR neutrino and NMC muon 
      structure functions $\frac{5}{6} F_2^{CCFR}-3F_2^{NMC}$ 
      is shown as solid 
      triangles. The strange quark distribution extracted 
      by the CCFR in a LO-analysis (Ref. \protect\cite{CCFRLO}) 
      is shown as solid stars, that from a NLO-analysis 
      (Ref.\ \protect\cite{CCFRNLO}) is represented by the 
      solid line with a band indicating  $\pm 1\sigma$ 
      uncertainty in the distribution.
      Statistical and systematic errors are added in
      quadrature. }
\label{fig4}
\end{figure}

In Fig.\ \ref{fig4} we show the results obtained for $x s(x)$ (open 
circles) and $x \bar s(x)$ (solid circles) by solving the resulting
linear equations, Eqs.\ \ref{diff} and \ref{s2}.  The results are
completely unphysical, since the extracted anti strange 
quark distribution is negative.  Our analysis strongly suggests that 
requiring charge symmetry, but allowing $s(x) \ne \bar{s}(x)$, 
cannot resolve the discrepancy between $F_2^{CCFR}(x,Q^2)$ and 
$F_2^{NMC}(x,Q^2)$.  The experimental results are incompatible,  
even if $\bar{s}(x)$ is completely unconstrained \cite{Brocm}. 

As neither neutrino shadowing corrections nor allowing $s(x) \ne 
\bar{s}(x)$ removes the low-$x$ discrepancy between the 
neutrino and muon structure functions, there remain two possible  
explanations.  Either one of the experimental 
structure functions (or the strange quark distribution) is incorrect 
at low $x$, or parton charge symmetry is violated in this
region.   
Assuming the possibility of parton CSV, we 
can combine the dimuon data for the strange quark distribution 
(Eq.\ \ref{s2}) with the relation between neutrino and muon
structure functions, Eq.\ \ref{diff} to obtain  
\begin{eqnarray} 
   \frac{5}{6} F_2^{CCFR}(x,Q^2) &-& 3 F_2^{NMC}(x,Q^2)
 -x s^{\mu\mu}(x) = {x(2\alpha -1)\over 3}[ s(x) -\bar{s}(x)] 
 \nonumber \\ &+& \frac{x}{6} \, [ 
  (5\alpha -1)(\delta d(x)-\delta u(x))+ 
  (4-5\alpha )(\delta\bar d(x) -\delta\bar u(x))] \nonumber \\ 
  &\approx & {x(2\alpha -1)\over 3}[s(x) 
 -\bar{s}(x)] + \frac{1}{2} \, x \, [\delta \bar d(x) -\delta 
\bar u(x)]. 
\label{csv1} 
\end{eqnarray} 
In Eq.\ \ref{csv1} we have used the experimental value 
$\alpha = \alpha^\prime$.  Since the experimental discrepancy is 
primarily in the small $x$-region, where sea quark
distributions are much larger than valence quarks, CSV effects should 
appear predominantly 
in the sea quark distributions. Setting  
$\delta q_v(x)=\delta q(x) -\delta \bar q(x) \approx 0$ in 
this region gives the second relation in 
Eq.(\ref{csv1}).  

\begin{figure}
\epsfig{figure=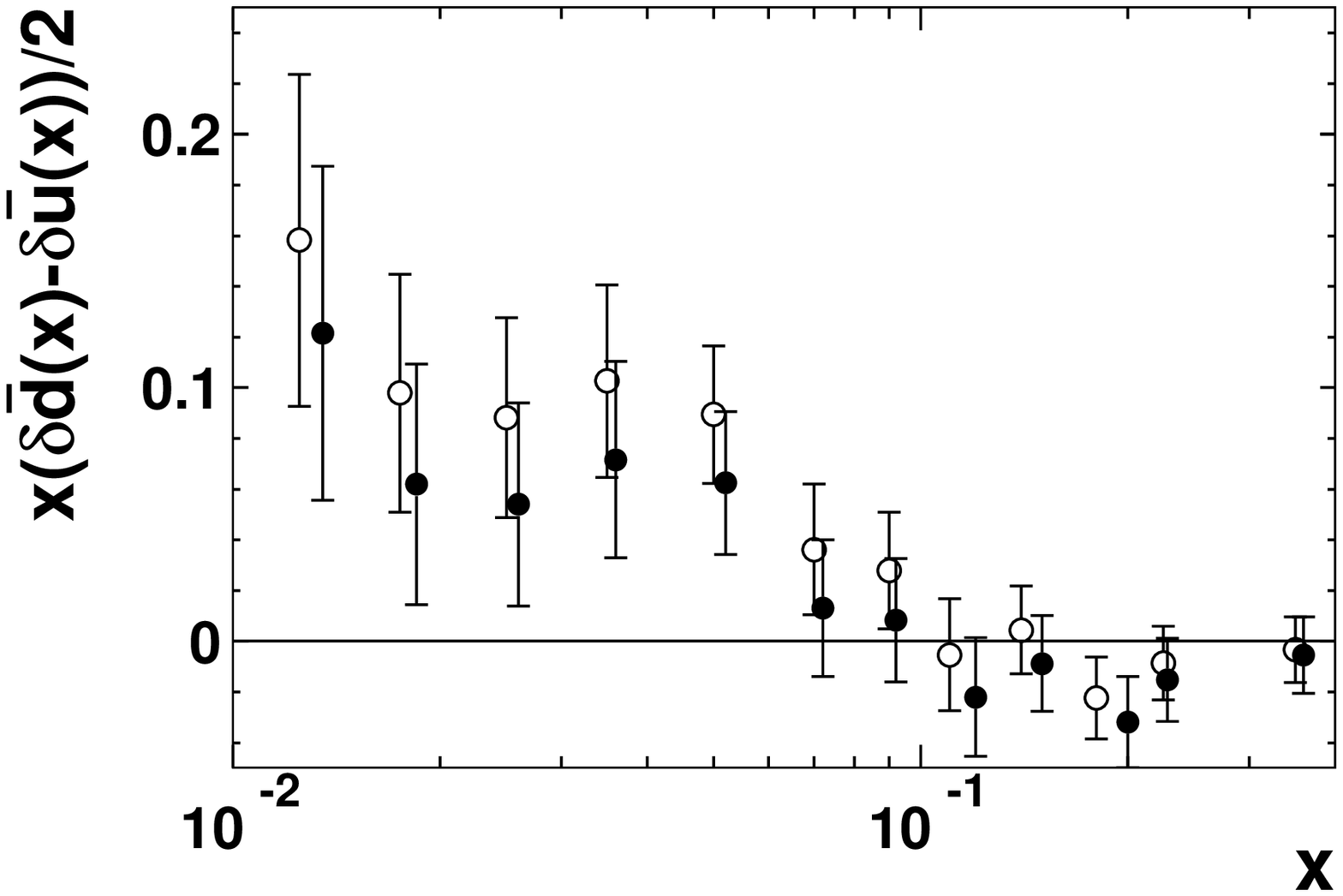,height=14.cm}
\caption{Charge symmetry violating distributions extracted 
      from the CCFR and NMC structure function data and 
      the CCFR dimuon production data under the assumption  
      that $s(x)=\bar s(x)$ (open circles) and 
      $\bar s(x)\approx 0$ (solid circles). Statistical and 
      systematic errors are added in quadrature. } 
\label{fig5}
\end{figure}

The left hand side of Eq.\ \ref{csv1} is positive.  Consequently, 
the smallest CSV effects will be obtained by setting   
$\bar{s}(x) = 0$.  In Fig.\ref{fig5} we show the 
magnitude of charge symmetry violation needed to satisfy the
experimental values in Eq.\ \ref{csv1}.  The solid circles 
are obtained if we set $\bar{s}(x) = 0$, and the open circles 
result from setting $\bar{s}(x) = s(x)$.  
Both the structure functions and the dimuon data  
have been integrated over the overlapping kinematical regions.  
In averaging the dimuon data we used the CTEQ4L parametrization for  
$s^{\mu\mu}(x)$ \cite{Lai}, and we extracted the strange quark 
distributions according to Eq.\ \ref{s2}. The CSV effect required 
to account for the NMC-CCFR discrepancy is extraordinarily large.  
It is roughly the same size as the strange quark distribution at
small $x$.  This CSV term is roughly 25\% of the 
light sea quark distributions for $x < 0.1$.  
The existing experimental data thus appears to require a very 
surprising, and uncomfortably large, violation of parton charge 
symmetry at small $x$.  

Clearly, CSV effects of this magnitude need further experimental 
verification.  It is hard to imagine how such large CSV effects are 
compatible with the high precision of charge symmetry measured 
at low energies.  The level of CSV required is 
at least two orders of magnitude larger than theoretical estimates 
of charge symmetry violation \cite{Ben98,Sather}.    
We will discuss the implications of such 
a large violation of charge symmetry in a subsequent paper \cite{Bor98}.  
Theoretical considerations suggest that $\delta\bar d(x) \approx 
-\delta\bar u(x)$ \cite{Ben98}. We note that with this sign CSV 
effects also require large flavor symmetry violation.  
At small $x$, our results can be summarized by 
\begin{eqnarray}
\delta \bar{d}(x) - \delta \bar{u}(x) &\approx & {1\over 4} 
  \left({\bar{u}(x) + \bar{d}(x) \over 2} \right) \approx {1\over 2} 
  (s(x) + \bar{s}(x) ) \nonumber \\ 
  \delta \bar{d}(x) + \delta \bar{u}(x) &\approx & 0 \quad\quad 
  (x < 0.1)
\label{csvappr}
\end{eqnarray} 
This suggests that for $x < 0.1$, $\bar{d}^n(x) \approx 
1.25\,\bar{u}^p(x)$ and $\bar{u}^n(x) \approx 0.75\,\bar{d}^p(x)$.  
If CSV effects of this magnitude are really present, then one must  
include charge symmetry violating quark distributions in phenomenological 
models from the outset, and re-analyze 
the extraction of all parton distributions.  

This work is supported in part by the Australian Research Council, 
and by the National Science Foundation under research contract
NSF-PHY9722706.  One of the authors [JTL] wishes to thank the
Special Research Centre for the Subatomic Structure of Matter for
its hospitality during the period when this work was carried out.

\references 

\bibitem{Miller} G. A. Miller, B. M. K. Nefkens and I. Slaus, Phys. Rep. 
       {\bf 194} (1990) 1; E. M. Henley and G. A. Miller in {\it Mesons 
         in Nuclei}, eds M. Rho and D. H. Wilkinson 
         (North-Holland, Amsterdam 1979). 

\bibitem{Lon98} J.T. Londergan and A.W. Thomas, to be published
          in {\it Progress in Particle and Nuclear Physics}, 1998.

\bibitem{NMCfsv} NMC-Collaboration, P. Amaudruz {\it et al.}, 
         Phys. Rev. Lett. {\bf 66} (1991) 2712; 
         Phys. Lett. {\bf B295} (1992) 159. 

\bibitem{Gottfried} K. Gottfried, Phys. Rev. Lett. {\bf 18} (1967) 1174. 

\bibitem{Na51} NA51-Collaboration, A. Baldit {\it et al.}, 
          {\em Phys. Lett.} {\bf B332} (1994) 244.  

\bibitem{E866} E866-Collaboration, E. A. Hawker  {\it et al.}, 
            to be published in Phys. Rev. Lett.  

\bibitem{Ma1} B.-Q. Ma, Phys. Lett. {\bf B274} (1992) 433;  
    B.-Q. Ma, A. W. Sch\"afer and W. Greiner, Phys. Rev. 
             {\bf D47} (1993) 51. 

\bibitem{Tim1} J. T. Londergan, S. M. Braendler and A. W. Thomas, 
            Phys. Lett. {\bf B424}, 185 (1998); J. T. Londergan, Alex 
            Pang and A.W. Thomas, Phys. Rev {\bf D54} (1996) 3154. 

\bibitem{CCFR} CCFR-Collaboration, W.G.Seligman {\it et al.},
{\em Phys. Rev. Lett.} {\bf 79}, 1213 (1997) and W.G.Seligman Ph.D. Thesis,
Nevis Report 292.

\bibitem{NMC} NMC-Collaboration, M.Arneodo {\it et al.}, {\em Nucl. Phys.}
{\bf B483}, 3 (1997).

\bibitem{Boros} C. Boros, J. T. Londergan and A. W. Thomas, 
       (preprint hep-ph/9804411). 

\bibitem{Badelek} J. Kwiecinski and B. Badelek, {\em  Phys. Lett.}
 {\bf B208}, 508 (1988); W. Melnitchouk and A.W. Thomas, {\em Phys. Lett.}
 {\bf B317}, 437 (1993); Phys.\ Rev. {\bf C52}, 3373 (1995).

\bibitem{Melni97} W. Melnitchouk and M. Malheiro,
        {\em Phys. Rev.} {\bf C55}, 431 (1997); 
	X. Ji and J. Tang, 
	{\em Phys. Lett.} {\bf B362}, 182 (1995); 
	H. Holtmann, A. Szczurek and J. Speth,
        {\em Nucl. Phys.} {\bf A596}, 631 (1996).

\bibitem{CCFRLO} S.A.Rabinowitz {\it et al.}, CCFR-Collaboration,
{\em Phys. Rev. Lett.} {\bf 70}, 134 (1993).

\bibitem{CCFRNLO} CCFR-Collaboration, A.O. Bazarko {\it et al.},
{\em Z. Phys.} {\bf C65},
189 (1995).

\bibitem{Bor98} C. Boros, J.T. Londergan and A.W. Thomas, to be 
published.

\bibitem{Brocm} In Ref.\protect\cite{Brodsky} it was suggested that allowing 
$s(x)\ne \bar s(x)$ could account for the difference between the two 
determinations of the strange quark distribution.  
This result was obtained by assuming  
$\alpha=\frac{1}{2}$ in Eq.\ \protect\ref{diff} and $\alpha^\prime =1$ in 
Eq.\ \protect\ref{s2}.  The choice $\alpha=\frac{1}{2}$ does not agree with
experiment ($\alpha = 0.83$). 

\bibitem{Brodsky} S.J.Brodsky and B.Q.Ma, {\em Phys. Lett.}
 {\bf B381}, 317 (1996).

\bibitem{Lai}  H. L. Lai {\it et al.}, {\em Phys. Rev.} {\bf D55}, 
   1280 (1997)

\bibitem{Ben98} E. Rodionov, A. W. Thomas and J. T. Londergan,
              {\em Mod. Phys. Lett.} {\bf A9}, 1799 (1994); 
              C.J. Benesh and J.T. Londergan, preprint
	      {\it nucl-th}/9803017.

\bibitem{Sather} E. Sather, {\it Phys. Lett.} {\bf B 274}, 433 (1992); 
	C.J. Benesh and T. Goldman, Phys.\ Rev.\ {\bf C55},
	441 (1997).
\end{document}